\documentclass[twocolumn,showpacs,preprintnumbers,prd,amssymb]{revtex4}

\begin{document}
\preprint{IFT.P-85/2002}
\title{Teleparallel origin of the Fierz picture for spin-2 particle}
\author{Yu.N. Obukhov}
\altaffiliation[Permanent address: ]{Department of 
Theoretical Physics, Moscow State University, 117234 Moscow, Russia}
\author{J.G. Pereira}
\affiliation{Instituto de F\'{\i}sica Te\'orica,
Universidade Estadual Paulista\\
Rua Pamplona 145,
01405-900 S\~ao Paulo SP, Brazil}

\begin{abstract}
A new approach to the description of spin-2 particle in flat and curved
spacetime  is developed on the basis of the teleparallel gravity theory. 
We show  that such an approach is in fact a true and natural framework 
for the Fierz representation proposed recently by Novello and Neves. More
specifically, we demonstrate how the teleparallel theory fixes uniquely
the structure of the Fierz tensor, discover the transparent origin of
the gauge symmetry of the spin 2 model, and derive the linearized Einstein 
operator from the fundamental identity of the teleparallel gravity. In 
order to cope with the consistency problem on the curved spacetime, 
similarly to the usual Riemannian approach, one needs to include the 
non-minimal (torsion dependent) coupling terms.  
\end{abstract}

\maketitle

\section{Introduction}

A consistent description of higher (greater than 1) spin fields interacting
with external classical (electromagnetic and gravitational) fields
represents a nontrivial problem which is not completely solved until now.
Earlier analyses of this problem can be found in \cite{FP,ara}, and a more
recent discussion, as well as a list of basic references, is given in
\cite{buch}. Recently, a new approach based on the so called Fierz theory 
of the spin 2 particle in flat and curved spacetime has been developed by
Novello and Neves \cite{novello}. In this work, using Fierz variables, the
consistency problem of a massive spin 2 field in a curved spacetime has been
studied, with the results indicating that the Fierz representation seems to
be more appropriate to deal with the coupling of a spin-2 field to gravity.

Now, as is well known, curvature and torsion are able to provide each one
equivalent descriptions of the gravitational interaction. In fact, according
to general relativity, curvature is used to {\it geometrize} spacetime, and
in this way successfully describe the gravitational interaction.
Teleparallelism, on the other hand, attributes gravitation to torsion, but 
in this case torsion accounts for gravitation not by geometrizing the
interaction, but by acting as a {\it force}. This means that, in the
teleparallel equivalent of general relativity, there are no geodesics, but
force equations quite analogous to the Lorentz force equation of
electrodynamics~\cite{sp1}. Gravitational interaction, thus, can be described
{\em alternatively} in terms of curvature, as is usually done in general
relativity, or in terms of torsion, in which case we have the so called
teleparallel gravity.
 
An important point of teleparallel gravity is that it corresponds to a
gauge theory for the translation group. Due to the peculiar character of
translations, any gauge theory including these transformations will differ
from the usual internal gauge models in many ways, the most significant 
being the presence of a tetrad field. On the other hand, a tetrad field can
naturally be used to define a linear Weitzenb\"ock connection, which is a
connection presenting torsion, but no curvature. A tetrad field can also
be used to define a Riemannian metric, in terms of which a Levi-Civita
connection is constructed. As is well known, it is a connection presenting 
curvature, but no torsion. It is important to keep in mind that torsion and 
curvature are properties of a connection~\cite{livro}, and many different 
connections can be defined on the same manifold. Therefore one can say that 
the presence of a nontrivial tetrad induces both a teleparallel and a
Riemannian structures in spacetime. The first is related to the
Weitzenb\"ock, and the second to the Levi-Civita connection.

The reason for gravitation to present two equivalent descriptions is related
to a quite peculiar property of gravitation, the so called {\em universality}.
Let us explore this point in more details. Like any other interaction,
gravitation presents a description in terms of a gauge theory. In fact, as
mentioned above, teleparallel gravity is a gauge theory for the translation
group, with contortion playing the role of force. On the other hand,
universality of gravitation means that all particles feel gravity the same.
In other words, particles with different masses feel a different gravitational
force in such a way that all these particles acquire the same acceleration. As
a consequence of this property, it turns out to be possible to describe
gravitation not as a {\em force}, but as a {\em deformation} of spacetime.
More precisely, according to this view, a gravitational field is supposed to
produce a {\em curvature} in spacetime, the gravitational interaction being
achieved in this case by letting test particles to follow the geodesics of
spacetime. This is the approach used by the general relativity description of
gravitation. It is important to notice that only an interaction presenting the
universality property can be described by a {\em geometrization} of spacetime.

Instead of using the general relativity approach, the basic purpose of this
paper will be to use the teleparallel approach to analyze the problem of a
spin-2 field coupled to gravity. The teleparallel approach will be used to
describe both the spin-2 field, and the gravitational background field.
The basic conclusion will be that the Fierz picture constructed in
\cite{novello} is naturally present in the teleparallel construction. In this
sense, we can say that the teleparallel equivalent of general relativity
appears to be a natural framework to deal with the spin-2 theory. In fact,
the small perturbations of the tetrad field are shown to reproduce correctly
the behavior of the spin-2 field on the flat Minkowski spacetime. The
antisymmetric piece of the tetrad turns out to be redundant, although taking
into account its explicit contribution makes the underlying gauge symmetry
more transparent. The generalization for the presence of gravitation is
straightforward, and it represents an alternative way to describe the spin-2
particle interacting with an external gravitational field.

\section{Teleparallel gravity: basic facts}

The general structure of teleparallel gravity is presented in detail in
\cite{sp1,HS,sp2,sp3,PR}. In this section we summarize the fundamentals of this
theory. In short, teleparallel approach can be understood as a gauge theory
of the spacetime translation group. Using the Greek alphabet to denote
spacetime indices, and the Latin alphabet to denote the local frame
components, the corresponding gauge potential is represented by the
nontrivial part of the tetrad field $h^{a}{}_{\mu}$. This tetrad gives rise
to the so called Weitzenb\"ock connection,
\begin{equation}\label{carco}
\Gamma^{\rho}{}_{\mu\nu} = h_{a}{}^{\rho}\partial_{\nu}h^{a}{}_{\mu},
\end{equation}
which introduces the distant parallelism on a four-dimensional spacetime
manifold. It is a connection that presents torsion, but not curvature. Its
torsion,
\begin{equation}
T^{\rho}{}_{\mu\nu} = \Gamma^{\rho}{}_{\nu\mu} - 
\Gamma^{\rho}{}_{\mu\nu}, \label{tor}
\end{equation}
plays the role of the translational gauge field strength. The Weitzenb\"ock
connection can be conveniently decomposed into the Riemannian and the
post-Riemannian pieces,
\begin{equation}
\Gamma^{\rho}{}_{\mu\nu} = \widetilde{\Gamma}^{\rho}{}_{\mu\nu} 
+ K^{\rho}{}_{\mu\nu},
\end{equation}
where
\[
\widetilde{\Gamma}^{\rho}{}_{\mu\nu} = {\frac 12}g^{\rho\sigma}
\left(\partial_\mu g_{\nu\sigma} + \partial_\nu g_{\mu\sigma} 
- \partial_\sigma g_{\mu\nu}\right)
\]
is the Christoffel symbol constructed from the spacetime metric $g_{\mu\nu} =
h^a{}_\mu h^b{}_\nu
\eta_{ab}$, and
\begin{equation}
K^{\rho}{}_{\mu \nu} = {\textstyle \frac{1}{2}} \left( 
T_{\mu}{}^{\rho}{}_{\nu} + T_{\nu}{}^{\rho}{}_{\mu} 
- T^{\rho}{}_{\mu \nu} \right)
\end{equation}
is the contortion tensor. Correspondingly, all other geometrical and physical
objects and operations constructed with the help of the Riemannian
connection $\widetilde{\Gamma}^{\rho}{}_{\mu\nu}$ will be denote
with a tilde. 

The gauge gravitational field Lagrangian reads
\begin{equation}\label{gala}
{\cal L}_G = \frac{c^{4}}{16\pi G}\,h\,S^{\rho\mu\nu}\,T_{\rho\mu\nu},
\end{equation}
where $h = {\rm det}(h^{a}{}_{\mu})$, and
\begin{equation}
S^{\rho\mu\nu} = - S^{\rho\nu\mu} := {\textstyle \frac{1}{2}} 
\left[ K^{\mu\nu\rho} - g^{\rho\nu}\,T^{\sigma\mu}{}_{\sigma} 
+ g^{\rho\mu}\,T^{\sigma\nu}{}_{\sigma} \right]
\label{S}
\end{equation}
is a tensor written in terms of the Weitzenb\"ock connection only.
Inverting this equation, we obtain torsion in terms of the above tensor:
\begin{equation}
T^{\rho}{}_{\mu\nu} = 2(S_\mu{}^\rho{}_\nu - S_\nu{}^\rho{}_\mu)
+ \delta^\rho_\mu\,S^\sigma{}_{\nu\sigma} - \delta^\rho_\nu
\,S^\sigma{}_{\mu\sigma}.  
\end{equation}
The Lagrangian (\ref{gala}) describes what is commonly known as the
teleparallel equivalent of Einstein's general relativity theory.
Performing a variation with respect to the tetrad, we find the teleparallel
version of the gravitational field equation,
\begin{equation}
\partial_\sigma(h S_\lambda{}^{\sigma \rho}) -
\frac{4 \pi G}{c^4} \, (h t_\lambda{}^\rho) = 0,
\label{eqs1}
\end{equation}
where
\begin{equation}
h t_\lambda{}^\rho = \frac{c^4 h}{4 \pi G} \, S_{\mu}{}^{\nu \rho}
\,\Gamma^\mu{}_{\nu\lambda} - \delta_\lambda{}^\rho \, {\cal L}_G
\label{emt}
\end{equation}
is the energy-momentum (pseudo) tensor of the gravitational field. This
equation is known to be equivalent to the Einstein's equation of general
relativity. It is important to notice that the left-hand side of the field
equation (\ref{eqs1}) can be rewritten as the usual left-hand side
of Einstein equations
\begin{equation}
\partial_\sigma(h S_\lambda{}^{\sigma \rho}) - \frac{4 \pi G}{c^4} 
\,(h t_\lambda{}^\rho) \equiv {\frac h2}\left(\widetilde{R}_\lambda{}^\rho 
- {\frac 12}\,\delta_\lambda^\rho\,\widetilde{R}\right),\label{ident}
\end{equation}
which then provides an easy proof of the Lemma 2 of \cite{novello}. As 
the source of both field equations is the symmetric energy-momentum tensor, 
the equivalence alluded to above holds also in the presence of matter 
\cite{blago}. It is worth noticing that the teleparallel field equation 
has the same structure of the Yang-Mills equation, which is consistent 
with the fact that teleparallel gravity corresponds to a gauge theory. 
We see in this way that the teleparallel approach to gravitation is more 
closely related to field theory than the general relativity approach.

\section{Linearized theory}

The trivial tetrad $h^a{}{}_\mu = \delta^a{}_\mu$ describes the flat geometry,
for which the metric has the diagonal Minkowski form, $g_{\mu\nu} =
\eta_{\mu\nu} = {\rm diag}(+1, -1, -1, -1)$. Let us then expand the tetrad
field around the flat background as follows,
\begin{equation}
h^a{}_\mu = \delta^a{}_\mu + u^a{}_\mu.\label{hexp1}
\end{equation}
The Weitzenb\"ock connection reads, correspondingly,
\begin{equation}
\Gamma^{\rho}{}_{\mu\nu} = \partial_{\nu}\,u^{\rho}{}_{\mu},\label{conL}
\end{equation}
where $u^\rho{}_\mu = \delta_a{}^\rho\,u^a{}_\mu$. As a result, the 
torsion and its trace are, respectively,
\begin{equation}
T^{\rho}{}_{\mu\nu} = \partial_{\mu}\,u^{\rho}{}_{\nu} 
- \partial_{\nu}\,u^{\rho}{}_{\mu},\quad 
T^{\rho}{}_{\mu\rho} = \partial_{\mu}\,u 
- \partial_{\rho}\,u^{\rho}{}_{\mu},
\end{equation}
with $u = u^\rho{}_\rho$. Decomposing the perturbation tensor $u_{\mu \nu}$
into the symmetric and antisymmetric pieces,
\begin{equation}
u_{\mu\nu} = \phi_{\mu\nu} + a_{\mu\nu},\label{upa1}
\end{equation}
with
\begin{equation}
\phi_{\mu\nu} := u_{(\mu\nu)} \quad {\rm and} \quad a_{\mu\nu} :=
u_{[\mu\nu]}, \label{upa2}
\end{equation}
we compute immediately the contortion tensor:
\begin{equation}
K^{\rho}{}_{\mu\nu} = \partial^{\rho}\,\phi_{\mu\nu} 
- \partial_{\mu}\,\phi^{\rho}{}_{\nu} + \partial_{\nu}\,a^{\rho}{}_{\mu}.
\end{equation}

Substituting now the above expressions into (\ref{S}), we obtain
\begin{eqnarray}
S^{\rho\mu\nu} = {\frac 1 2} &\Big[& \partial^{\mu}\,\phi^{\nu\rho} 
- \partial^{\nu}\,\phi^{\mu\rho} - g^{\rho\nu}\left(\partial^\mu\,\phi
- \partial_{\sigma}\,\phi^{\sigma\mu}\right) \nonumber \\
&+& g^{\rho\mu}\left(
\partial^\nu\,\phi - \partial_{\sigma}\,\phi^{\sigma\nu}\right)
+ \,\partial^{\rho}\,a^{\mu\nu} \nonumber \\
&+& g^{\rho\nu}\,\partial_{\sigma}\,
a^{\sigma\mu} - g^{\rho\mu}\,\partial_{\sigma}\,a^{\sigma\nu}\Big].
\label{S1}
\end{eqnarray}
Comparison with \cite{novello} shows that the first line is nothing but
the Fierz tensor $F^{\mu\nu\rho} = -\,F^{\nu\mu\rho}$ introduced 
by No\-vel\-lo and Neves. More specifically:
\begin{equation}
S^{\rho\mu\nu} = -\,F^{\mu\nu\rho} + {\frac 1 2}\left(\partial^{\rho}
\,a^{\mu\nu} + g^{\rho\nu}\,\partial_{\sigma}\,a^{\sigma\mu} 
- g^{\rho\mu}\,\partial_{\sigma}\,a^{\sigma\nu}\right). 
\end{equation}
Since we have identically
\[
\partial_\mu\left(\partial^{\rho}
\,a^{\mu\nu} + g^{\rho\nu}\,\partial_{\sigma}\,a^{\sigma\mu} 
- g^{\rho\mu}\,\partial_{\sigma}\,a^{\sigma\nu}\right)\equiv 0,
\]
the last term drops out completely from the linearized gravitational
field equations (\ref{eqs1}), which then reads:
\begin{equation}
\partial_\sigma\,S_\lambda{}^{\sigma\rho} = 0.\label{eqs2}
\end{equation}
This equation yields the correct dynamics of the spin 2 particle in
flat spacetime, as it is easily seen from the identity (\ref{ident}).
Indeed, substituting the expansion (\ref{hexp1}) into it, we find that 
\begin{equation}
\partial_\sigma\,S_\lambda{}^{\sigma\rho}\equiv {\frac 12}
\,\widetilde{G}^{\rm L}{}_\lambda{}^\rho,
\end{equation}
where the left-hand side represents the linearized Einstein tensor:
\begin{eqnarray}
\widetilde{G}^{\rm L}_{\mu\nu} &=& \Box\left(\eta_{\mu\nu}\,\phi - \phi_{\mu\nu}
\right) - \partial_\mu\partial_\nu\,\phi \nonumber \\
&-& \eta_{\mu\nu}\,\partial_\alpha
\partial_\beta\,\phi^{\alpha\beta}
+ \partial_\mu\partial_\lambda
\,\phi^\lambda{}_\nu + \partial_\nu\partial_\lambda\,\phi^\lambda{}_\mu.
\end{eqnarray}
The identity (\ref{ident}) plays the fundamental role in the teleparallel
theory, since it underlies the proof of the equivalence of Einstein's
gravity and the teleparallel gravity. Now, as a by-product of this identity
we have straightforwardly derived the Lemma 2 of \cite{novello}.

It is interesting to notice that the teleparallel Lagrangian (\ref{gala}) 
can be rewritten as
\begin{equation}
{\cal L}_G = \frac{c^{4}}{8\pi G}\,h\left(S^{\rho\mu\nu}\,S_{\rho\mu\nu}
- S^{\sigma\mu}{}_\sigma\,S^\rho{}_{\mu\rho} - 3\,S^{\rho\mu\nu}
\,S_{[\rho\mu\nu]}\right).\label{gala2}
\end{equation}
Inserting here (\ref{S1}), we can verify that the antisymmetric field 
$a_{\mu\nu}$ drops out completely, in accordance with the analysis of 
the linearized field equations. This observation shows that, as a matter of
fact, the antisymmetric field does not have physical importance. Indeed, one
can show quite generally that, by means of a local Lorentz transformation,
it is always possible to choose a frame in which the tetrad matrix is
symmetric \cite{dewitt,ogi,ruben}. In such a frame, the field $a_{\mu\nu}$
vanishes, and as a result the tensor $S^{\rho\mu\nu}$ coincides  (up to a
sign) with the Fierz tensor introduced in \cite{novello}.

However, there is a certain reason to keep $a_{\mu\nu}$ nontrivial. In
particular, we can verify the covariance of the linearized formalism with
respect to ``general coordinate'' spacetime transformations. The
demonstration of this property in \cite{novello} is rather long and
not very transparent. In contrast, here it is sufficient to notice that the
tensor (\ref{S1}) is explicitly {\it invariant} under the gauge
transformations:
\begin{eqnarray}
\phi_{\mu\nu} &\longrightarrow& \phi_{\mu\nu} + \partial_\mu\,\Lambda_\nu
+ \partial_\nu\,\Lambda_\mu,\label{gaugeP}\\
a_{\mu\nu} &\longrightarrow& a_{\mu\nu} - \partial_\mu\,\Lambda_\nu
+ \partial_\nu\,\Lambda_\mu.\label{gaugeA}
\end{eqnarray}
The proof is straightforward: One just needs to substitute these formulas
into Eq. (\ref{S1}). The original perturbation field (\ref{upa1}) transforms 
as $u^\mu{}_\nu \longrightarrow u^\mu{}_\nu + 2 \, \partial_\nu \Lambda^\mu$, 
in complete agreement with the geometrical meaning of the tetrad.

\section{Spin-2 field in the presence of gravitation}

In order to construct the theory for a spin-2 particle on a curved
spacetime, instead of the expansion (\ref{hexp1}), we consider the
tetrad expansion
\begin{equation}
h^a{}_\mu = \overline{h}{}^a{}_\mu + u^a{}_\mu
\label{hexp2}
\end{equation}
around the nontrivial classical background $\overline{h}{}^a{}_\mu$. Such
an approach is analogous to the treatment of a consistent spin-2 model as 
a first order perturbation of the general relativity theory \cite{ara}. We 
will denote with an overline every other background objects and operations.
In order to simplify the computations, in contrast to the above described
flat-space discussion, we will choose the symmetric gauge
\cite{dewitt,ogi,ruben} from the very beginning. Then, we have the symmetric
tensor field 
\begin{equation}
\phi^\mu{}_\nu := \overline{h}_a{}^\mu\,u^a{}_\nu; \quad 
\phi_{\mu\nu} = \phi_{\nu\mu},
\end{equation}
where $\overline{h}_a{}^\mu$ is the inverse background tetrad:
$\overline{h}_a{}^\mu \,\overline{h}{}^a{}_\nu = \delta^\mu_\nu$. From now
on, the Greek indices will be raised and lowered with the help of the
background metric $\overline{g}_{\mu\nu} =\overline{h}{}^a{}_\mu
\,\overline{h}{}^b{}_\nu \, \eta_{ab}$.  

Substituting the expansion (\ref{hexp2}) into the Weitzenb\"ock connection
(\ref{carco}), up to first order in $\phi^\rho{}_\mu$, we find
\begin{equation}
\Gamma^{\rho}{}_{\mu\nu} = \overline{\Gamma}^{\rho}{}_{\mu\nu} + 
\overline{\nabla}_\nu \phi^\rho{}_\mu, \label{con1}
\end{equation}
where $\overline{\Gamma}^{\rho}{}_{\mu\nu} = \overline{h}_{a}{}^{\rho}
\partial_{\nu}\overline{h}{}^{a}{}_{\mu}$ is the background teleparallel 
connection, and $\overline{\nabla}_\nu$ is the covariant derivative in the
connection $\overline{\Gamma}^{\rho}{}_{\mu\nu}$. As a result, we obtain 
for the torsion 
\begin{equation}
T^{\rho}{}_{\mu\nu} = \overline{T}^{\rho}{}_{\mu\nu} + \overline{\nabla}_\mu
\phi^\rho{}_\nu - \overline{\nabla}_\nu\phi^\rho{}_\mu.\label{tor1}
\end{equation}
Using this expression in Eqs. (\ref{S}) and (\ref{gala2}), we
straightforwardly obtain the kinetic term for the spin 2 field Lagrangian in
the presence of gravitation,
\begin{eqnarray} \label{gala3}
{\cal L}_G &=& \frac{c^{4}}{8\pi G}\,\overline{h}\left(F^{\rho\mu\nu}
\,F_{\rho\mu\nu} - F^{\sigma\mu}{}_\sigma\,F^\rho{}_{\mu\rho}\right) \nonumber \\
&=& \frac{c^{4}}{8\pi G}\,\overline{h}\,F^{\rho\mu\nu}\,\overline{\nabla}_\nu
\phi_{\mu\rho}.
\end{eqnarray}
Here, $F^{\rho\mu\nu}$ is the covariant generalization of the Fierz tensor:
\begin{eqnarray}
F^{\rho\mu\nu} = {\frac 1 2}\Big[\overline{\nabla}^{\nu}\phi^{\mu\rho} 
&-& \overline{\nabla}^{\mu}\phi^{\nu\rho} + g^{\rho\nu}\left(
\overline{\nabla}^\mu\phi - \overline{\nabla}_{\sigma}\phi^{\sigma\mu}
\right) \nonumber \\
&-& g^{\rho\mu}\left(\overline{\nabla}^\nu\phi - 
\overline{\nabla}_{\sigma}\phi^{\sigma\nu}\right)\Big].\label{Fcurv}
\end{eqnarray}

The main difference of the spin-2 model (\ref{gala3}) from the theory
studied in \cite{novello} is that the covariant derivatives in (\ref{Fcurv})
are defined with the help of the teleparallel connection, and not with the
help of the Riemannian (Christoffel) connection.  We thus obtain an
alternative way to describe a spin 2 particle in the presence of gravitation. 
The resulting theory is obviously generally covariant.

As it is well known \cite{FP,ara,buch,novello}, the higher spin theories
are generally inconsistent in the presence of the electromagnetic and/or
gravitational field. Technically, this amounts to the non-vanishing 
covariant divergence of the corresponding field operator which, in turn,
is related to the fact that the covariant derivative has a nontrivial
commutator proportional to the curvature. The consistency conditions, 
derived for the higher spin fields, place strong restrictions on the
spacetime curvature which are not fulfilled, in general. 

In the teleparallel gravity, the spacetime curvature is zero, whereas 
the commutator of the covariant derivatives reads
\begin{equation}
\left[\nabla_\mu, \nabla_\nu\right] = -\,T^\lambda{}_{\mu\nu}\,\nabla_\lambda.
\end{equation}
Accordingly, the consistency condition for the theory (\ref{gala3}) will
be not algebraic, as in the usual formulation, but differential instead:
$\overline{T}^\lambda{}_{\mu\nu}\,\overline{\nabla}_\lambda F^{\rho\mu\nu} =0$.

If we demand that the consistency conditions should be satisfied for all
field configurations, we then have to conclude that the torsion should
vanish. This observation agrees with the general analysis of the higher
spin theories on the Riemann-Cartan spacetime \cite{high}. A possible 
way to avoid the consistency problem is to include non-minimal 
coupling terms in the Lagrangian \cite{ara,buch}. Since the curvature
is zero, the non-minimal terms may involve only the torsion tensor. The
latter, being of the 3rd rank, necessarily involves the derivative for 
the construction of the invariant contractions. It is straightforward
to see that, adding the non-minimal interaction Lagrangian
\begin{eqnarray}
{\cal L}_{non} &=& \frac{c^{4}}{8\pi G}\,\overline{h}\Big[
\phi_{\alpha\lambda}\left(K^\lambda{}_{\mu\nu} - T^\lambda{}_{\mu\nu}
\right)\widetilde{\nabla}^\nu\,\phi^{\mu\alpha} \nonumber \\
&-& \phi^{\alpha\mu}
\,T_{\lambda\mu\nu}\,\widetilde{\nabla}_\alpha\,\phi^{\lambda\nu}
- \,\phi^{\lambda\nu}\,T_{\lambda\mu\nu}
\left(\widetilde{\nabla}^\mu\,\phi - \widetilde{\nabla}_\alpha
\,\phi^{\alpha\mu}\right) \nonumber \\
&+& \phi^{\mu\nu}\,T^\lambda{}_{\mu\lambda}
\left(\widetilde{\nabla}_\nu\,\phi - \widetilde{\nabla}^\alpha
\,\phi_{\alpha\nu}\right)\Big]\label{nonmin}
\end{eqnarray}
to the Lagrangian (\ref{gala3}) yields the Riemannian Fierz Lagrangian
of \cite{novello}. This allows then to solve the consistency problem.

The theory of massive spin-2 particle \cite{ara,buch} was satisfactorily
formulated within the framework of the Fierz approach \cite{novello}.
Our results also admit a direct generalization to the nontrivial mass
which we though do not discuss here since it follows along the same
lines as in \cite{novello}.

\section{Discussion and conclusions}

Novello and Neves \cite{novello} have demonstrated that the Fierz 
representation for the spin 2 theory has a number of advantages as
compared to the alternative approaches (such as the non-ambiguity
of the order of derivatives and the equivalence to the non-minimal
curvature Einstein representation with the fixed coefficients of additional
terms \cite{ara}). Here, we have shown that the Fierz representation
can be naturally understood on the basis of the teleparallel gravity.
In particular, we find that (i) the structure of the Fierz tensor defined 
in \cite{novello} through an {\it ad hoc} procedure is unambiguously fixed by
the  teleparallel theory; (ii) the gauge symmetry (\ref{gaugeP})-(\ref{gaugeA})
underlying the corresponding spin 2 freedom, is manifested straightforwardly;
(iii) the linearized Einstein operator arises immediately as a consequence
of the fundamental identity (\ref{ident}) of the teleparallel theory. 
It is worthwhile to note that the antisymmetric piece of the tetrad 
field, while being dynamically redundant, plays a significant role 
in the formal derivations. A certain disadvantage of our approach 
is the need of the non-minimal coupling of the type (\ref{nonmin})
to solve the consistency problem on the curved spacetime. This is 
similar to the observations made within the Riemannian approach \cite{ara}.

Summarizing, in this paper we have developed a new approach to the
description of spin 2 in flat and curved spacetime on the basis of
the teleparallel gravity theory. This approach appears to be a true 
origin for the Fierz representation proposed recently in \cite{novello}.

\begin{acknowledgments}
The work of YNO was supported by FAPESP. JGP thanks FAPESP and CNPq for
partial financial support. We are grateful to Friedrich Hehl for
valuable advices.
\end{acknowledgments}

\end{document}